\documentclass[conference]{IEEEtran}
\usepackage{cite}
\usepackage[pdftex]{graphicx}
\usepackage{dblfloatfix}
\usepackage{array}
\usepackage[hyperfootnotes=false]{hyperref}

\newcommand\blfootnote[1]{%
  \begingroup
  \renewcommand\thefootnote{}\footnote{#1}
  \addtocounter{footnote}{-1}%
  \endgroup
}

\begin{document}

\title{Voice Pathology Detection Using Deep\\Learning: a~Preliminary Study}
\author{
	\IEEEauthorblockN{
		Pavol Harar\IEEEauthorrefmark{1},
		Jesus B. Alonso-Hernandez\IEEEauthorrefmark{2},
		Jiri Mekyska\IEEEauthorrefmark{1},
		Zoltan Galaz\IEEEauthorrefmark{1},
		Radim Burget\IEEEauthorrefmark{1} and
		Zdenek Smekal\IEEEauthorrefmark{1}
	}
	\IEEEauthorblockA{
		\IEEEauthorrefmark{1}
			Department of Telecommunications, Brno University of Technology, Technicka~10, 61600~Brno, Czech Republic
	}
	\IEEEauthorblockA{
		\IEEEauthorrefmark{2}
		Institute for Technological Development and Innovation in Communications (IDeTIC)\\
		University of Las Palmas de Gran Canaria, 35001 Las Palmas de Gran Canaria, Spain
	}
}

\maketitle

\begin{abstract}
This paper describes a~preliminary investigation of Voice Pathology Detection using Deep Neural Networks (DNN). We used voice recordings of sustained vowel /a/ produced at normal pitch from German corpus Saarbruecken Voice Database (SVD). This corpus contains voice recordings and electroglottograph signals of more than 2\,000 speakers. The idea behind this experiment is the use of convolutional layers in combination with recurrent Long-Short-Term-Memory (LSTM) layers on raw audio signal. Each recording was split into 64\,ms Hamming windowed segments with 30\,ms overlap. Our trained model achieved 71.36\,\% accuracy with 65.04\,\% sensitivity and 77.67\,\% specificity on 206 validation files and 68.08\,\% accuracy with 66.75\,\% sensitivity and 77.89\,\% specificity on 874 testing files. This is a~promising result in favor of this approach because it is comparable to similar previously published experiment that used different methodology. Further investigation is needed to achieve the state-of-the-art results.

\end{abstract}

\IEEEpeerreviewmaketitle

\section{Introduction}

\vspace{-0.2cm}
\blfootnote{978-1-5386-0850-0/17/\$31\copyright2017 IEEE. Personal use of this material is permitted. Permission from IEEE must be obtained for all other uses, in any current or future media, including reprinting/republishing this material for advertising or promotional purposes, creating new collective works, for resale or redistribution to servers or lists, or reuse of any copyrighted component of this work in other works. \href{https://doi.org/10.1109/IWOBI.2017.7985525}{DOI: 10.1109/IWOBI.2017.7985525}.\\
Email: \href{mailto:harar@phd.feec.vutbr.cz}{harar@phd.feec.vutbr.cz}}

According to~\cite{Mekyska2015} the automatic detection of vocal fold pathologies is a~task of assigning normophonic or dysphonic labels to a~given phonation produced by a~specific speaker. This objective is an interest to the researchers of speech or voice community, as well as the respective medical community. This is due to its non-invasive nature, free from subjective biasness, and relatively low cost. So far, many researchers aimed to detect voice pathology by analyzing the voice with the emphasis to develop features that can effectively distinguish between normal and pathological voices~\cite{Muhammad2017A}.

On the contrary, in this paper we investigate a~way to skip the phase of developing the features. Instead, we aim to create an end-to-end deep neural network model capable of voice pathology assessment using raw audio signal. To achieve this goal, we used voice recordings from Saarbruecken Voice Database (SVD)~\cite{Woldert2007} that contains the samples of healthy persons and patients with one up to 71 different pathologies.

Nowadays, thanks to huge increases in computational power and data amounts, the Deep Learning (DL) models delivered the state-of-the-art results in many domains including Speech processing. Using this approach to tackle the voice pathology detection problem we are allowed to use complex multi-layer model architectures. We expect the convolutional layers~\cite{LeCun1998} to learn to detect various patterns that could help us to differentiate between healthy and pathological voice. Long-Short-Term-Mermoy layers~\cite{Hochreiter1997} should then transform the time distributed abstract feature vectors outputted from convolution stacks into understandable representation for fully connected dense layers, which should do the final classification.

The rest of this paper is organized as follows. Section \ref{sec:related} introduces the related works in this area of expertise. In Section \ref{sec:methodology}, data and methodology of the experiment are be discussed. The results are presented in Section \ref{sec:results}. Conclusions are drawn in Section \ref{sec:conclusions}.

\section{Related Work}
\label{sec:related}

There is already a~great number of related works in this area of expertise \cite{Muhammad2017A, Alnasheri2014, Martinez2012, Souissi2015, Souissi2016, Alnasheri2017A, Hossain2016, Eskidere2015, Alnasheri2017B, Muhammad2017B, Hemmerling2016}.

Detailed information about papers published on SVD can be found in Table\,\ref{tab:relatedWork}. In summary, the authors that used SVD extracted various features from the voice recordings prior to pathology detection. The features were usually extracted from time, frequency and cepstral domains and contained mel-frequency cepstral coefficients (MFCC), energy, entropy, short-term cepstral parameters, harmonics-to-noise ratio, normalized noise energy, glottal-to-noise excitation ratio, multidimensional voice program parameters (MDVP), etc. After the feature extraction, multiple classifiers have been used. Most authors relied on Support Vector Machines (SVM) and Gaussian Mixture Models (GMM) but K-means clustering (KM), Random forests (RF), Extreme Learning Machines (ELM) and Artificial Neural Networks (ANN) were also utilized in several papers. To our best knowledge, this is the first paper that presents the voice pathology detection using DNN.

\begin{table*}[htb!]
	\caption{Overview of related works}
	\label{tab:relatedWork}
	\centering
	
	\begin{tabular}{ l  p{5.5cm}  p{3cm}  l  p{5.5cm} }
		\hline
		\hline
		Article & Feature set & Employed classifier & Accuracy & Notes\\
		\hline
		
			\cite{Hemmerling2016} & 28 parameters extracted from time, frequency and cepstral domain & KM, RF & 100.00\,\% & Used combination of vowels /a/, /i/, /u/ \break Females and Males separately \\
			\cite{Muhammad2017B} & energy, entropy, contrast, homogeneity & GMM & 99.98\,\% & Used combination of voice and EGG signals \\
			\cite{Alnasheri2017B} & MDVP parameters & SVM & 99.68\,\% & Used subset containing 4 of 71 pathologies \\
			\cite{Eskidere2015} & MFCC & GMM & 99.00\,\% & Used combination of vowels /a/, /i/, /u/ \\
			\cite{Hossain2016} & MPEG-7 low-level audio and IDP & SVM, ELM, GMM & 95.00\,\% & Used mix of MEEI and SVD data \\
			\cite{Muhammad2017A} & IDP & SVM & 93.20\,\% & Used subset containing 3 of 71 pathologies \\
			\cite{Alnasheri2017A} & Maximum peak and lag & SVM & 90.98\,\% & Used subset containing 4 of 71 pathologies \\
			\cite{Souissi2016} & MFCC first and second derivatives & ANN & 87.82\,\% & Used subset containing 4 of 71 pathologies \\
			\cite{Souissi2015} & short-term cepstral parameters & SVM & 86.44\,\% & Used subset containing 4 of 71 pathologies \\
			\cite{Martinez2012} & MFCC, harmonics-to-noise ratio, normalized noise energy, glottal-to-noise excitation ratio & GMM & 79.40\,\% & Used combination of vowels /a/, /i/, /u/ \\
			\cite{Alnasheri2014} & Peak value and lag for every frequency band & GMM, SVM & 72.00\,\% & Used 200 samples of vowel /a/ at high pitch \\
		
		\hline
		\hline	
	\end{tabular}
\end{table*}

The results vary greatly between the published papers mainly due to differences between sets of data that were used for the experiment. Mart\'{i}nez et al. in~\cite{Martinez2012} reported 72\,\% accuracy using 200 recordings of sustained vowel /a/ at high pitch, which is the most similar experiment to ours. All other authors used combination of vowels /a/, /i/ and /u/. Souissi et al. in~\cite{Souissi2015, Souissi2016} reported the highest accuracy of 87.82\,\% using a~subset containing 4 types of pathologies from the total number of 71 as well as Al-nasheri et al. in~\cite{Alnasheri2017A, Alnasheri2017B} who pushed the accuracy of 99.68\,\%. The reason to use a~subset containing only some of the pathologies was to conduct an experiment on data that were also present in other available databases, namely Massachusetts Eye and Ear Infirmary Database (MEEI) and Arabic Voice Pathology Database (AVPD). Muhammad et al. in~\cite{Muhammad2017A} used subset containing 3 types of pathology and reported 93.20\,\% accuracy and then in~\cite{Muhammad2017B} he used combination of voice recordings as well as electroglottograph (EGG) signals to boost the accuracy to 99.98\,\%. The highest possible accuracy of 100\,\% was achieved by Hemmerling et al. in~\cite{Hemmerling2016} who approached the detection problem separately for female and male speakers. However, since the accuracy is so high the reported results are questionable.

\section{Methodology}
\label{sec:methodology}

\subsection{Data}
\label{sec:data}

We used Saarbruecken Voice Database, which is a~collection of voice recordings and EGG signals from more than 2\,000 persons. It contains recordings of 687 healthy persons (428 females and 259 males) and 1356 patients (727 females and 629 males) with one or more of the 71 different pathologies. One recording session contains the recordings of the following components:
\begin{itemize}
\item Vowels /i/, /a/, /u/ produced at normal, high and low pitch
\item Vowels /i/, /a/, /u/ with rising-falling pitch
\item Sentence ``Guten Morgen, wie geht es Ihnen?'' (``Good morning, how are you?'')
\end{itemize}

All samples of the sustained vowels are between 1 and 3 seconds long, sampled at 50\,kHz with 16-bit resolution~\cite{Woldert2007}. In contrary to MEEI database, all audio samples (healthy and pathological) in SVD were recorded in the same environment. This preliminary experiment was conducted using samples of sustained vowel /a/ produced at normal pitch. Each file was split into multiple 64\,ms long segments (Hamming windowed) with 30\,ms overlap. One file was therefore represented to the input of the neural network as a~matrix containing total number of $n$ (segments) $\cdot$ 3\,200 features (0.064\,s $\cdot$ 50\,000\,Hz = 3\,200 features).

We divided all data into TRAIN (70\,\%), VALIDATION (15\,\%) and TESTING (15\,\%) sets and we assured that the number of healthy and pathological samples in training and validation sets are equal. The rest was appended to the testing set. In total, there were 960 samples (480 healthy, 480 pathological) in the training set, 206 samples (103 healthy, 103 pathological) in validation set and the rest 874 samples (104 healthy and 770 pathological) were used as testing samples.

\subsection{DNN Architecture}
\label{sec:architecture}

While constructing the network, it is always good to have a~clear ``story'' in mind that would reason the task of every layer or stack of layers in the proposed architecture. The ``story'' behind our architecture is simple. We used 2 stacks of convolutional layers to transform the input vectors into a set of more abstract repeating patterns that seem important for the network cost to decrease. Between each stack of convolutions, there is a~pooling layer~\cite{Hinton2012} that reduces the dimensionality of the vector. Since each file is a~sequence of multiple time-steps (segments), all convolutions and pooling layers were wrapped in TimeDistributed layer (built in layer in Keras framework~\cite{Chollet2015} for keeping the time axis unchanged). Afterwards we reshaped the resulting matrices from the last pooling layer so it could be connected to the recurrent LSTM layer. Before the experiment, we legitimized the presence of LSTM to ourselves as a~context learning element that remembers the changes in time. As the last component of our network, there is a~stack of 3 fully connected layers ended with Softmax layer with 2 neurons (one neuron for class = healthy and the other neuron for class = pathological) for the final classification.

For the first two convolutional layers we used 16 kernels of size 160 succeeded with max pooling layer of size 4. The second stack of another two convolutional layers used 13 kernels of size 320 again succeeded with the same max pooling as before. Then we connected the flattened output from the last layer to the LSTM layer with 25 units. To prevent overfitting we set the dropout probability~\cite{Srivastava2014} on LSTM layer to 0.1 for input gates and 0.5 for the recurrent connections. From this point on the DNN used only fully connected layers. The first two of size 32 and the last one with 2 output neurons and Softmax activation~\cite{Bridle1990}.

\begin{figure}[t!]
	\centering
		\includegraphics[width=0.327\textwidth]{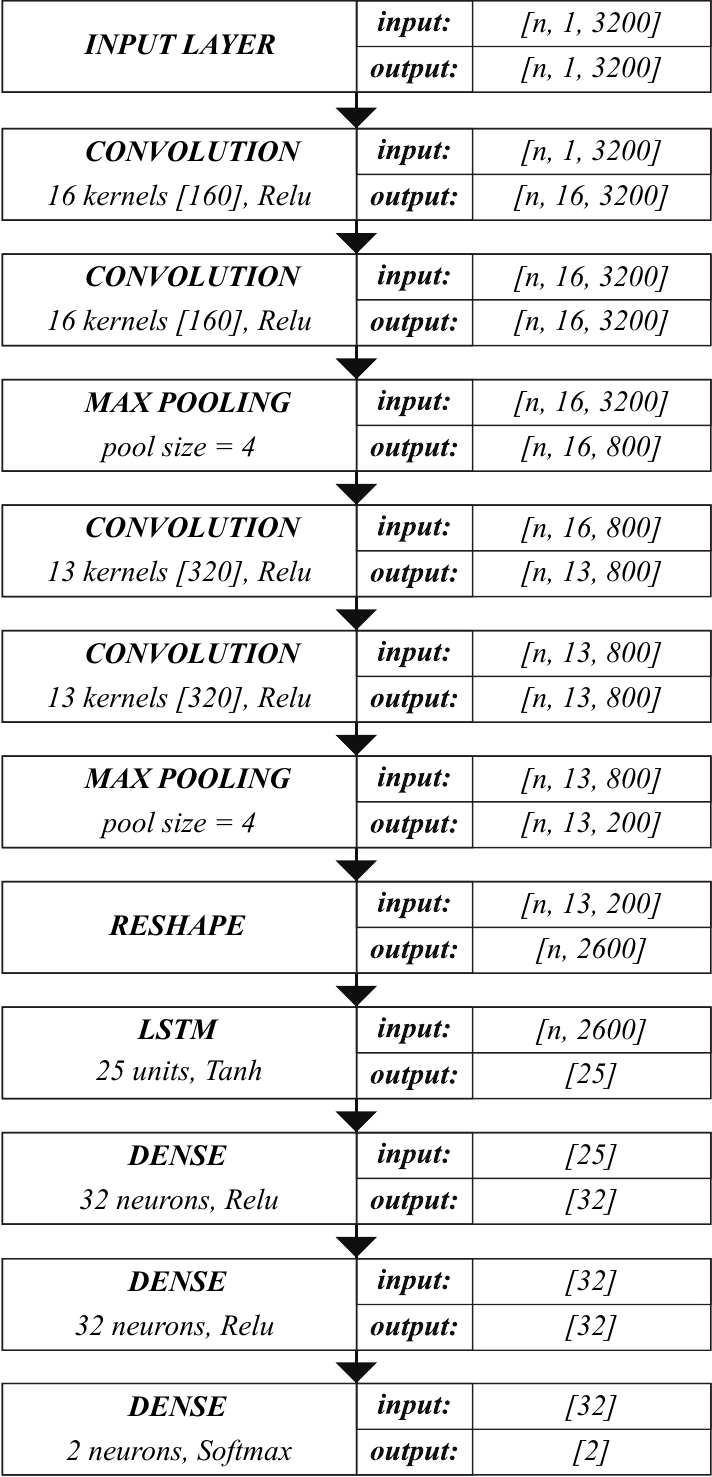}
	\caption{Detailed DNN architecture.}
	\label{fig:layers}
\end{figure}

Rectified linear unit (Relu) as activation function~\cite{Nair2010} was used for all convolutional and dense layers except the Softmax output layer. LSTM used Hyperbolic tangent (Tanh) activation function. All layers were initialized using Glorot uniform initialization~\cite{Glorot2010}. This whole DNN had overall 428\,772 trainable parameters and its whole architecture is depicted in Fig.\,\ref{fig:layers}.

DNN models consist of manifold of hyper-parameters. Sensitivity of each model on particular hyper-parameter is different due to a distinctive nature of the system that is modeled. Our strategy of its selection and fine tuning with aim of finding the best performing model was based on community standards, intuition and grid search for which we have utilized our open-source library KEX\footnote{KEX avaliable from http://splab.cz/en/download/software/kex-library}. 

\subsection{Experimental Setup}
\label{sec:setup}

For gradient-based optimization of Cross-entropy loss function used during training of our proposed model we utilized Adam algorithm~\cite{Kingma2014} with initial learning rate of $6 \cdot 10^{-5}$. The learning rate was not fixed and was decreased by factor 0.5 each time there was no improvement in validation accuracy for 8 consecutive training epochs (iterations). The minimum learning rate was set to $1 \cdot 10^{-7}$.

The data were presented to the DNN one file at a~time (batch size = 1) in a~matrix of size $n$~(the number of segments)~$\cdot$~3\,200~features for 34 epochs. We chose to use batch size equal to 1 because the length of each file is different, therefore each of the matrices had different number of segments. If we wanted to make the batches bigger, we would have to either put together files of the exact same length or cut the files to the same length.

To eliminate unnecessary training we set the patience equal to 20. That means the experiment was terminated if no progress on validation loss had been made for more than 15 epochs of training. The best results were recorded after the 25\textsuperscript{th} epoch. In order to train the DNN on GPU (Nvidia GeForce GTX 690) and build the models quickly, we utilized the capabilities of Keras framework. The whole 25 epochs long training took 101 minutes to finish. All hyper-parameters were tuned based on validation results.

\section{Results}
\label{sec:results}

In order to perform a~pathology detection using voice signal, we built a~deep neural network model consisting of convolutional, pooling, LSTM and fully connected layers. We trained, validated and tested it using recordings of sustained vowel /a/ produced at normal pitch from Saarbruecken Voice Database containing 71 types of pathologies. The signal was split into 64\,ms long Hamming windowed segments with 30\,ms overlap and was presented to the neural network as a~sequence of vectors in time. The training and validation sets contained exactly the same number of healthy and pathological samples as can be seen in Tab.\,\ref{tab:valicon}.

Out of 206 validation samples, the proposed trained model predicted 59 samples to belong to a~wrong class as opposed to 147 correct predictions resulting in 71.36\,\% validation accuracy with 65.04\,\% sensitivity (recall of class pathological) and 77.67\,\% specificity (recall of class healthy). The precision, recall and f1-score of validation samples is shown in Tab.\,\ref{tab:valicla}.

Tab.\,\ref{tab:testcon}. shows the DNN predicted 279 testing samples to belong to a~wrong class as opposed to 595 correct predictions  resulting in 68.08\,\% testing accuracy with 66.75\,\% sensitivity and 77.89\,\% specificity. The precision, recall and f1-score of validation samples is shown in Tab.\,\ref{tab:testcla}.

\begin{table}[htb!]
	\caption{VALIDATION confusion matrix}
	\label{tab:valicon}
	\centering
	
	\begin{tabular}{l c c c}
		\hline
		\hline
			& true: pathological & true: healthy & no. of segments \\
		\hline

			pred: pathological & 67 & 36 & 103\\
			pred: healthy & 23 & 80 & 103\\

		\hline
	\end{tabular}
\end{table}

\begin{table}[htb!]
	\caption{VALIDATION classification report}
	\label{tab:valicla}
	\centering
	
	\begin{tabular}{p{4.3cm} c c c}
		\hline
		\hline
		class & precision & f1-score & recall \\
		\hline
	
			pathological & 0.74 & 0.69 & 0.65\\
			healthy & 0.69 & 0.73 & 0.78\\
			
			\hline
			\multicolumn{3}{l}{overall accuracy:} & 71.36\,\% \\
		
		\hline
	\end{tabular}
\end{table}

\begin{table}[htb!]
	\caption{TESTING confusion matrix}
	\label{tab:testcon}
	\centering
	\begin{tabular}{l c c c}
		\hline
		\hline
		& true: pathological & true: healthy & no. of segments \\
		\hline
			pred: pathological & 514 & 256 & 770\\
			pred: healthy & 23 & 81 & 104\\
			
		\hline
	\end{tabular}
\end{table}

\begin{table}[htb!]
	\caption{TESTING classification report}
	\label{tab:testcla}
	\centering
	
	\begin{tabular}{p{4.3cm} c c c}
		\hline
		\hline
		class & precision & f1-score & recall \\
		\hline
		
			pathological & 0.96 & 0.79 & 0.67\\
			healthy & 0.24 & 0.37 & 0.78\\
			
			\hline
			\multicolumn{3}{l}{overall accuracy:} & 68.08\,\% \\
			
		\hline
	\end{tabular}
\end{table}

\section{Conclusions}
\label{sec:conclusions}

The objective of this paper was to carry out a~preliminary study which would clarify whether the use of Deep Neural Network model, especially combination of convolutional and LSTM layers would prove itself worthy of further exploration in case of Voice Pathology Detection problem using only sustained vowel. Using just recordings of vowel /a/ produced at normal pitch, the examined method achieved 71.36\,\% accuracy on validation data and 68.08\,\% accuracy on testing data. Since this result is comparable to that published in~\cite{Alnasheri2014} we assume that further investigation is in place and could lead to much better results.

The main advantage of this approach is that one does not need to build the feature vector as opposed to the previously proposed methods, thus it saves great amount of time and expertise in the area of the problem being solved. On the other hand, the main disadvantage is the amount of data needed to train the model which is also a~limitation of this experiment. The SVD database is extensive in numbers of persons recorded, but there is not enough samples of healthy persons in comparison with the number of samples of pathological patients. Also the distribution of individual pathologies is extremely unequal making the Voice Pathology Detection a~hard problem, because some of the samples with certain type of pathology that occurs just once in the whole dataset can end up in testing set. Hence the network could not be trained to recognize it resulting in worse accuracy.

Our future work will build on current experiment, but we will limit the number of pathologies only to those having the most samples as in~\cite{Muhammad2017A, Alnasheri2017A, Souissi2015, Souissi2016} and we will train separate models for males and females as in~\cite{Hemmerling2016}. We will investigate whether training with combination of vowels /a/, /i/ and /u/ help to improve the accuracy as in~\cite{Martinez2012, Eskidere2015, Hemmerling2016}. Also we will incorporate the data from other publicly available datasets and introduce permutation test to validate if the model learned to recognize meaningful features or just overfits on noise or remembers the samples.

\section*{Acknowledgment}

This work was supported by the grant of the Czech Ministry of Health 16-30805A (Effects of non-invasive brain stimulation on hypokinetic dysarthria, micrographia, and brain plasticity in patients with Parkinsons disease) and the following projects: SIX (CZ.1.05/2.1.00/03.0072), and LOl401. For the research, infrastructure of the SIX Center was used.

\bibliographystyle{IEEEtran}
\bibliography{ms}
\end{document}